\documentclass{article}
\usepackage{spconf,amsmath,graphicx}
\usepackage{multirow}
\usepackage{graphicx}
\usepackage{subfigure}
\usepackage{subfig}
\usepackage{cite}
\usepackage{caption}
\title{Simplified Self-Attention for Transformer-based End-to-End Speech Recognition}
\name{Haoneng Luo$^1$, Shiliang Zhang$^2$, Ming Lei$^2$, Lei Xie$^{1*}$}
\address{
  $^1$ Audio, Speech and Language Processing Group (ASLP@NPU), School of Computer Science, \\ Northwestern Polytechnical University, Xi'an, China\\
  $^2$ Speech Lab, Alibaba DAMO Academy}

\begin{document}
%\ninept
%
\maketitle
\begin{abstract}
 Transformer models have been introduced into end-to-end speech recognition with state-of-the-art performance on various tasks owing to their superiority in modeling long-term dependencies. However, such improvements are usually obtained through the use of very large neural networks. Transformer models mainly include two submodules -- position-wise feedforward layers and self-attention (SAN) layers. In this paper, to reduce the model complexity while maintaining good performance, we propose a simplified self-attention (SSAN) layer which employs FSMN memory blocks instead of projection layers to form query and key vectors for transformer-based end-to-end speech recognition. We evaluate the SSAN-based and the conventional SAN-based transformers on the public AISHELL-1, internal 1000-hour and 20,000-hour large-scale Mandarin tasks. Results show that our proposed SSAN-based transformer model can achieve over 20\% reduction in model parameters and 6.7\% relative CER reduction on the AISHELL-1 task. With impressively 20\% parameter reduction, our model shows no loss of recognition performance on the 20,000-hour large-scale task.
\end{abstract}
\renewcommand{\thefootnote}{\fnsymbol{footnote}}
\footnotetext{* Lei Xie is the corresponding author.}
\section{Introduction}
\label{sec:intro}

Conventional hybrid automatic speech recognition (ASR) systems have three main components, acoustic model, pronunciation model and language model, trained separately with individual optimization targets~\cite{hinton2012deep, dahl2011context}. In recent years, there has been significant progress on end-to-end (E2E)~\cite{chiu2018state} automatic speech recognition (ASR) which aims to combine the three models into a single neural network with the purpose to significantly simplify the construction of an ASR system. At present, there are mainly three E2E frameworks: connectionist temporal classification (CTC)~\cite{graves2006connectionist, hannun2014deep, zhang2018acoustic}, attention-based models~\cite{chorowski2015attention, chan2016listen} and transducers \cite{graves2012sequence, graves2013speech, rao2017exploring}. These models treat ASR as a sequence-to-sequence task that directly learns speech-to-text mapping with a neural network. These models can be combined as well to further boost performance~\cite{zhang2020transformer, kim2017joint, tian2019self-attention}. In this paper, we focus on attention-based models, aiming at better performance with simplified model structure.

  A typical attention-based model can be divided into three main components -- encoder, attention and decoder. For ASR tasks, the encoder extracts high-level acoustic features from input speech as acoustic model; the decoder extracts language features and predicts output sequence as pronunciation model and language model; attention module learns alignment between acoustic and language features. There are several structures of attention-based models, such as Listen, Attend and Spell (LAS)~\cite{chan2016listen} and transformer. Transformer~\cite{vaswani2017attention} is a typical sequence-to-sequence model that has made significant progress on various NLP tasks, such as machine translation, natural language understanding and language modeling. Recently, the transformer models have been applied to speech recognition tasks with competitive performance~\cite{povey2018time, sperber2018self, dong2018speech, Shigeki2019Improving}. As an attention-based encoder-decoder model, the core of this model is self-attention network (SAN) layers which can model long context dependencies. Besides, transformer has no recurrence structure, which can be trained much faster with more parallelization than models with recurrent components as in LAS~\cite{chan2016listen}. The transformer model was further explored by structure modification and model/loss integration. In~\cite{you2019dfsmn}, augmented persistent memory was applied to obtain more information beyond the whole utterance context length for self-attention layer and achieved improved performances on ASR tasks. In~\cite{Shigeki2019Improving}, the authors integrate CTC with transformer for joint training and decoding, which leads to significant improvements in various ASR tasks.

  E2E models, including the transformer, have great potential to be deployed in edge devices with a relatively compact foot-print and simpler building pipeline. However, the transformer models achieve superior recognition performance through stacking of many SAN layers, resulting in substantial increase in model parameters and severe decoding latency. For example, reported in\cite{pham2019very}, 48 SAN encoder layers plus another 48 SAN decoder layers totally constitute to 252M model parameters. Hence some variants have been explored to simplify the transformer models. In \cite{sukhbaatar2019augmenting}, an all-attention layer was proposed to reduce model size, where the self-attention and position-wise feedforward layers were merged by augmenting the self-attention layers with persistent memory vectors. It has shown that the additional persistent memory block in the form of key-value vectors can store some global information so that the bulky feedforward layers can be removed. This method simplifies the structure of the transformer model dramatically with no loss of performance on a language modeling task. 
    
 In this paper, we propose a new approach to simplify the self-attention layer while maintaining the performance superiority of a transformer model in speech recognition.  Specifically, we explore a simplified self-attention network (SSAN) layer by introducing FSMN memory block. Our work is inspired by the recent advances of feedforward sequential memory networks~\cite{zhang2015feedforward}. FSMN can effectively model long-term context dependency using a simple and elegant non-recurrent structure, achieving reduced model size and competitive performance over recurrent neural networks on both acoustic modeling and language modeling tasks~\cite{zhang2015feedforward}. In detail, in this paper, for each self-attention layer, we propose to form key-query vectors by FSMN memory blocks instead of projection layers, and the self-attention input is directly assigned to the value vectors without extra computation. By this way, key-query vectors can effectively store context information and further help the self-attention layer to capture long-term context dependencies. Meanwhile, the number of model parameters can be substantially reduced. The efficacy of our approach has been proved by experiments on several ASR tasks. On the open AISHELL-1 task, we obtain 6.7\% relative improvement in CER and 21.7\% reduction in model parameters as compared with a competitive baseline transformer model. Moreover, experiments on internal 1000- and 20,000-hour large-scale tasks show that the proposed SSAN-based transformer can effectively reduce the model parameters by 20\% with no loss of ASR performance.
  
 % The remainder of this paper is organized as follows. Section 2 describe the Transformer, FSMN and SSAN. Section 3 presents our experimental results. The conclusions are given in Section 4.
 
 \begin{figure}[t]
  \centering
  \begin{minipage}[t]{1.0\linewidth}
  \centering
  \includegraphics[width=2.7in]{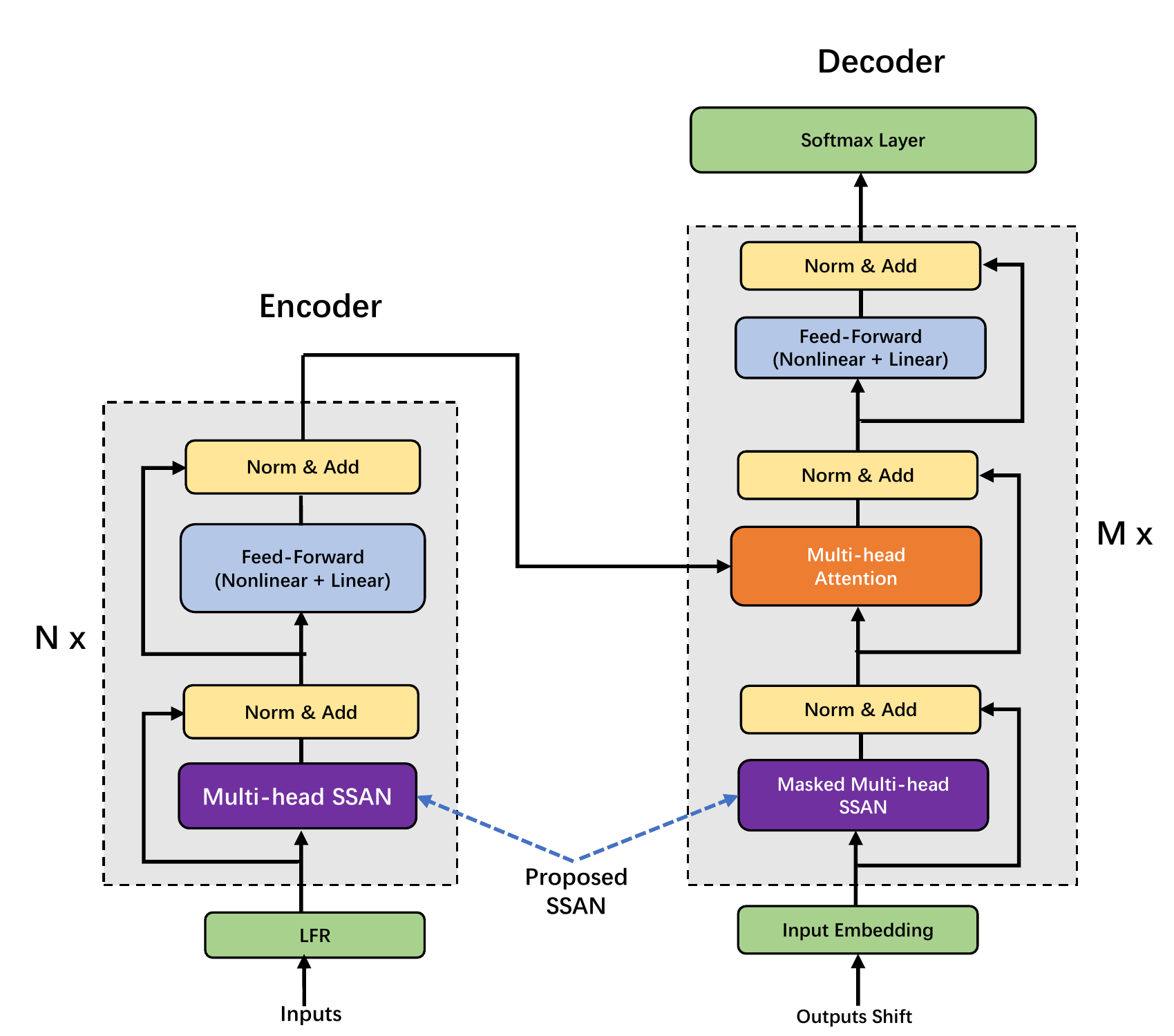}
  \end{minipage}%
  \caption{Illustration of the proposed SSAN-based Transformer. }
  \label{fig:ssan}
\end{figure}

\begin{figure*}[htb]
\centering
\subfigure[self-attention (SAN)]{
\begin{minipage}[t]{0.5\linewidth}
\centering
\includegraphics[width=2.7in]{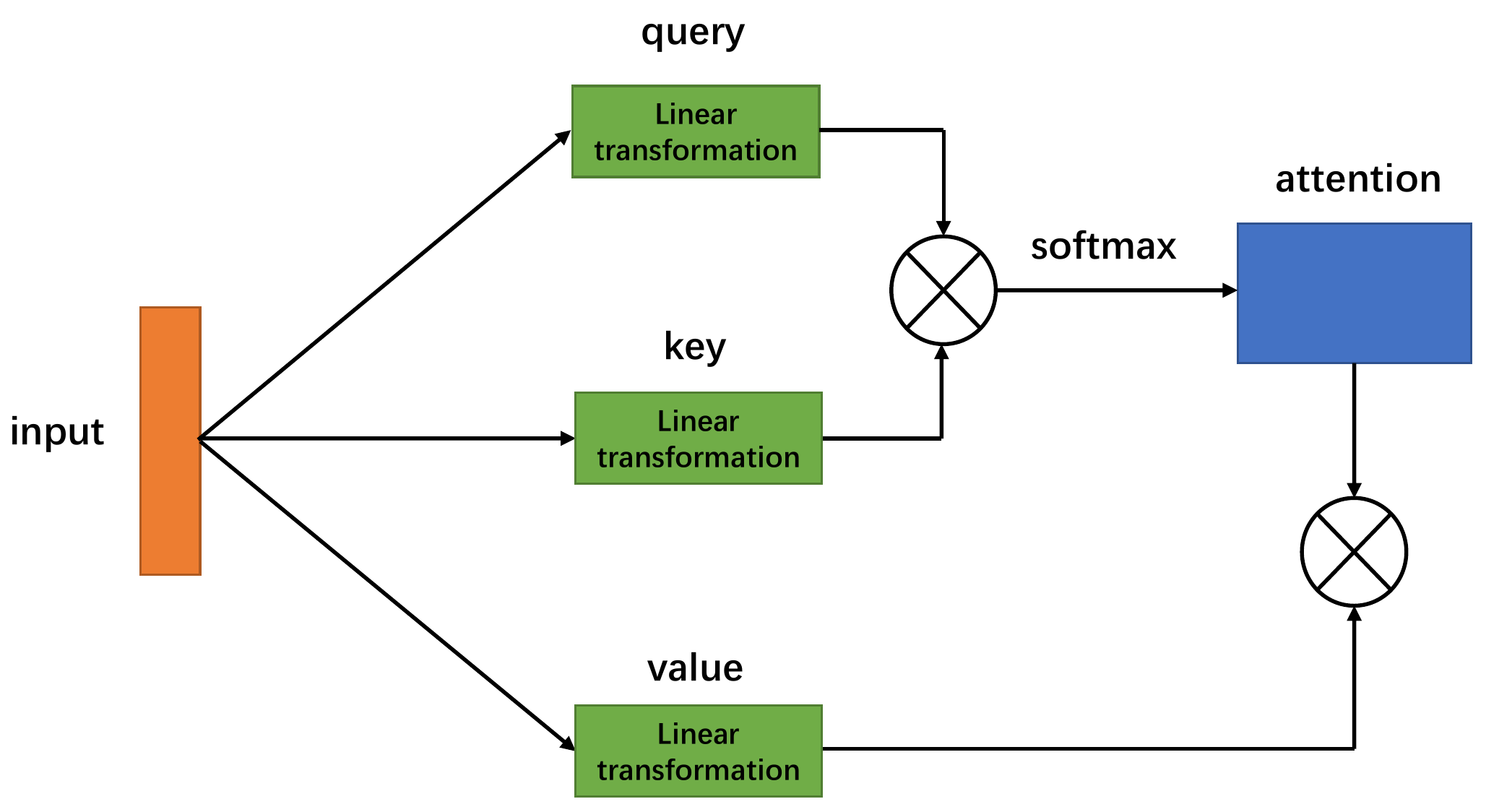}
%\caption{fig1}
\end{minipage}%
}%
\subfigure[simplified self-attention (SSAN)]{
\begin{minipage}[t]{0.5\linewidth}
\centering
\includegraphics[width=2.7in]{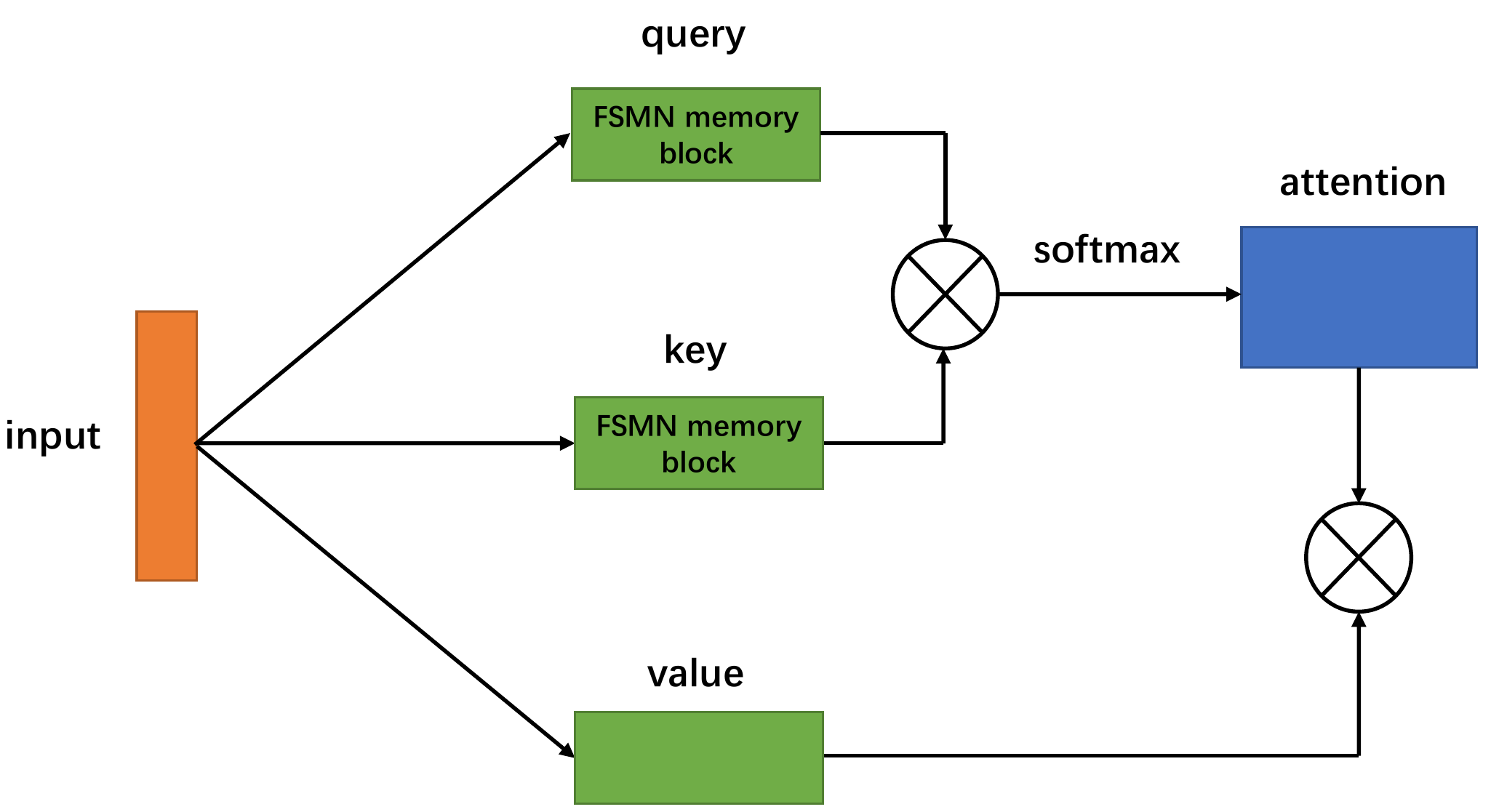}
%\caption{fig2}
\end{minipage}%
}%
\centering
\caption{In fig. 2(a), the query, key and value vector are formed by linear projection for self-attention network. In fig. 2(b), query and key vectors are formed by FSMN memory block, the input vector is directly assigned to value for simplified self-attention network.}
\label{fig:san-ssan}
\end{figure*}

\section{Model Architecture}
\label{sec:format}

As shown in Fig.~\ref{fig:ssan}, our modified SSAN-based transformer is built upon the typical transformer which has an attention-based encoder-decoder structure.  The encoder maps an input sequence of frame-level acoustic features $(x_1, ..., x_T)$ to a sequence of high-level representations $(h_1, ..., h_T)$ and the decoder generates a transcription $(y_1, ..., y_L)$ one token at a time step. The original self-attention network (SAN) based encoder has two sub-modules: a multi-head self-attention layer (encoder-attention) and a position-wise feedforward layer. The decoder network has three sub-modules including a masked multi-head self-attention layer (decoder-attention), a multi-head cross-attention layer between the encoder and the decoder, and a position-wise feedforward layer. Each layer is followed by a skip-connection and layer normalization.

  In the transformer~\cite{vaswani2017attention}, the input of each self-attention head is projected into query, key and value vectors. To simplify the self-attention layer, we replace the projection layer of querie and key vectors with FSMN memory block, while the input of self-attention is directly assigned to the value vectors. More details will be described in Section2.4. 
  
  As shown in Fig.~\ref{fig:ssan}, the original encoder and decoder self-attention layers are replaced by simplified self-attention network (SSAN) layers, while other modules remain unchanged.  

\subsection{Multi-head self-attention}
The core of a transformer model is multi-head self-attention layers which aim to capture long-term context dependencies. Multi-head self-attention is designed to jointly attend to information from different representation subspaces at different positions~\cite{vaswani2017attention}. Each attention head adopts the scaled dot-product attention to map a query and a set of key-value pairs to an output. The computation process of multi-head self-attention are formulated as follows.
\begin{equation}
  {\rm MultiHead}(Q, K, V) = {\rm Concat}({head_1}, ..., {head_h})W^O
  \label{eq4}
\end{equation}
\begin{equation}
  {head_i} = {\rm SelfAttn}(X{W_i}^Q, X{W_i}^K, X{W_i}^V)
  \label{eq3}
\end{equation}
\begin{equation}
  {\rm SelfAttn}(Q_i, K_i, V_i) = {\rm Softmax}(\frac{Q_{i}K_{i}}{\sqrt{d_k}})V_{i}
  \label{eq2}
\end{equation}
For ${head_i}$, ${W_i}^Q \in R^{d_{model} \times d_k}$, ${W_i}^K \in  R^{d_{model} \times d_k}$, ${W_i}^V \in R^{d_{model} \times d_v}$ are query, key and value projection matrices, respectively. ${W^O} \in R^{hd_{v} \times d_{model}}$ is the output projection matrix, ${h}$ denotes the number of heads, and $d_{model}$ is the attention dimension. In this work, we employ $d_{k} = d_{v} = d_{model} / h$.

\subsection{Position-wise Feedforward}
In addition to the multi-head self-attention layers, each layer in encoder and decoder contains a fully-connected feed-forward layer. This layer consists of two linear transformations with a ReLU activation in between:
\begin{equation}
  {\rm FFN}(X) = {\rm RELU}(XW_1 + b_1)W_2 + b_2
  \label{eq4}
\end{equation}
where $W_1$ and $W_2$ are matrices of dimension $d_{model} \times d_{ffn}$, and $b_1$ and $b_2$ are the bias.
\subsection{FSMN}\label{subsec:fsmn}
FSMN~\cite{zhang2015feedforward} extends the standard feedforward fully-connected neural networks by augmenting some memory blocks which function as FIR-like filters. The formulation of the memory block takes the following form:
\begin{equation}\label{eq.fsmn_eq}
  \bar{m}_t = m_{t} + \sum_{i=0}^{N_{1}} a_{i} \odot m_{t-i} + \sum_{j=1}^{N_{2}} c_{j} \odot m_{t+j},
\end{equation}
where $\odot$ denotes element-wise multiplication of two equally-sized vectors. $N_{1}$ is called the look-back order, denoting the number of historical items looking back to the past, and $N_{2}$ is called the look-ahead order, representing the size of the lookahead window into the future. From Eq.~(\ref{eq.fsmn_eq}), we can observe that the key element in FSMN is the learnable FIR-like filters, which are used to encode long-context information into fixed-size.

\subsection{The proposed SSAN}
Fig.~\ref{fig:san-ssan}(a) shows the self-attention layer form query, key, value by projection layers. Let's denote the input vector of self-attention layer as $X = [x_1, ..., x_T]$. The computation process of query, key, value is formulated as
\begin{equation}
  Q_t = W^{Q}x_{t},
  \label{eq5}
\end{equation}
\begin{equation}
  K_t = W^{K}x_{t},
  \label{eq6}
\end{equation}
and
\begin{equation}
  V_t = W^{V}x_{t},
  \label{eq7}
\end{equation}
respectively. In order to simplify the self-attention layer, we propose a new way to form query, key and value vectors, which is shown in Fig.~\ref{fig:san-ssan}(b). Specifically, we use the FSMN memory block introduced in Section~\ref{subsec:fsmn} to form query and key, while the input vector $X$ is directly assigned to value. Formally, query, key and value become
\begin{equation}
  Q_t = x_{t} + \sum_{i=0}^{N_{1}} a_{i} \odot x_{t-i} + \sum_{j=1}^{N_{2}} c_{j} \odot x_{t+j},
  \label{eq5}
\end{equation}
\begin{equation}
  K_t = x_{t} + \sum_{i=0}^{N_{1}} b_{i} \odot x_{t-i} + \sum_{j=1}^{N_{2}} d_{j} \odot x_{t+j},
  \label{eq6}
\end{equation}
and
\begin{equation}
  V_t = x_t
  \label{eq7}
\end{equation}
respectively. From the perspective of query, key and value formation, we can see that SAN itself considers no context information while the proposed SSAN can extract context information for calculation of the attention matrix due to the introduction of FSMN memory blocks.

As for model size, a SAN layer requires $3*d_{model}*d_{model}$ parameters to form query, key and value vectors while a SSAN layer requires $2*(N_1+N_2)*d_{model}$ parameters. If we take the FIR width $(N_1 + N_2)$ to be small, the number of parameters of SSAN will be much smaller than SAN.

\section{Experiments}
\label{sec:pagestyle}

\subsection{Dataset}
In this paper, we validated the proposed SSAN-based transformer on three Mandarin speech recognition datasets: public AISHELL-1 corpus~\cite{bu2017aishell}, internal 1000 and 20,000 hours corpus same as used in~\cite{zhang2018deep}. The AISHELL-1 corpus is composed of read speech from 400 speakers collected from high fidelity microphone, and the 20,000-hour corpus is collected from many service domains, such as sports, tourism, gaming, literature and others, which is more diverse in data and more challenging in speech recognition. The 1000-hour dataset is shuffled from the 20,000-hour corpus. For the AISHELL-1 task, we use the 150-hour training set for model training and the 10-hour development set for early-stopping. Finally, the character error rate (CER\%) is reported in the 7176-sentence test set (about 5 hours). As to the 1000/20,000-hour tasks, we use two types of test sets -- \emph{near-field} and \emph{far-field}. Far-field set consists of about 10 hours data and near-field set consists of about 5 hours data.
%The near-field set consists of 10 hours of speech from smartphones and desktop microphones while the more challenging far-field set consists of 5 hours of speech collected from distance (e.g., from a smart speaker) and enhanced by multi-channel signal front-end.

\subsection{Experimental Setup}
For all experiments, the input features are 80-dimensional log Mel-filterbank (FBank) computed on 25ms window with 10ms shift. We stack the consecutive frames within a context window of 7 (3-1-3) to produce the 560-dimensional FBank features and then downsample the input frame rate to 60ms. We also apply SpecAugment~\cite{park2019specaugment} for data augmentation. All the experiments are based on the transformer framework. We chose 4233 and 9000 characters (including \textless pad\textgreater, \textless eos\textgreater\ and \textless sos\textgreater\ labels) as model units for AISHELL-1, 1000/20,000-hour tasks respectively. All experiments are conducted using the open-source, sequence modeling toolkit -- OpenNMT \cite{Klein2017OpenNMT}.

  We employ $h$ = 8 parallel attention heads in the transform models. For every transformer layer, we use $d_k$ = $d_v$ = $d_{model}/{h}$ = 64, $d_{ffn}$ = 2048. For the FSMN memory block, we set $N_1=11$, $N_2=10$ for the encoder, and $N_1=11$, $N_2=0$ for the decoder. We adopt LazyAdamOptimizer\cite{vaswani2017attention} with $learning\_rate=1.0$, $warm\_up=8000$, and gradient clipping at 5.0. Moreover, we employ label smoothing and dropout regularization to prevent over-fitting.
  \begin{table}[!htb]
\normalsize
\caption{Results of different model architectures on AISHELL-1 test sets. SAN: self-attention network; SSAN: simplified self-attention network; \#L: the number of layers; M: Million.}  
\label{tab:diffstruc}  
\centering  
\begin{tabular}{|l|l|c|c|c}  
\hline  
Encoder (\#L) & Decoder (\#L) & Param. (M) & CER (\%) \\ \hline  
\hline
SAN (6) & SAN (3) & 34 & 7.75 \\ \hline  
SSAN (6) & SSAN (3) & 27 & \textbf{7.65} \\  \hline 
\hline
SAN (10) & SAN (3) & 46 & 7.33 \\ \hline  
SSAN (10) & SSAN (3) & 36 & \textbf{6.84} \\  \hline 
\hline
SAN (12) & SAN (6) & 64 & 7.97 \\ \hline  
SSAN (12) & SSAN (6) & 51 & \textbf{7.16} \\  \hline 
\end{tabular} 
\end {table}
\vspace{-0.4cm}
\begin{table}[!htb]
\normalsize
\caption{Comparison of SSAN and other published models on AISHELL-1.} 
\centering  
\begin{tabular}{|l|c|c|c}  
\hline  
Model & LM & CER (\%) \\ \hline \hline 
TDNN-LFMMI \cite{povey2016purely} & Y & 7.62 \\ \hline  
LAS \cite{shan2019component} & Y & 8.71 \\ \hline
Joint CTC-attention / ESPNet \cite{karita2019a} & Y & 6.70 \\ \hline
SSAN (ours) & N & \textbf{6.84} \\ \hline
\end{tabular}   
\label{tab:comparison}
\end {table}

\subsection{AISHELL-1 Task}
\begin{figure}[t]
\begin{minipage}[t]{0.33\linewidth}
\centering
\centerline{\includegraphics[width=1.1in]{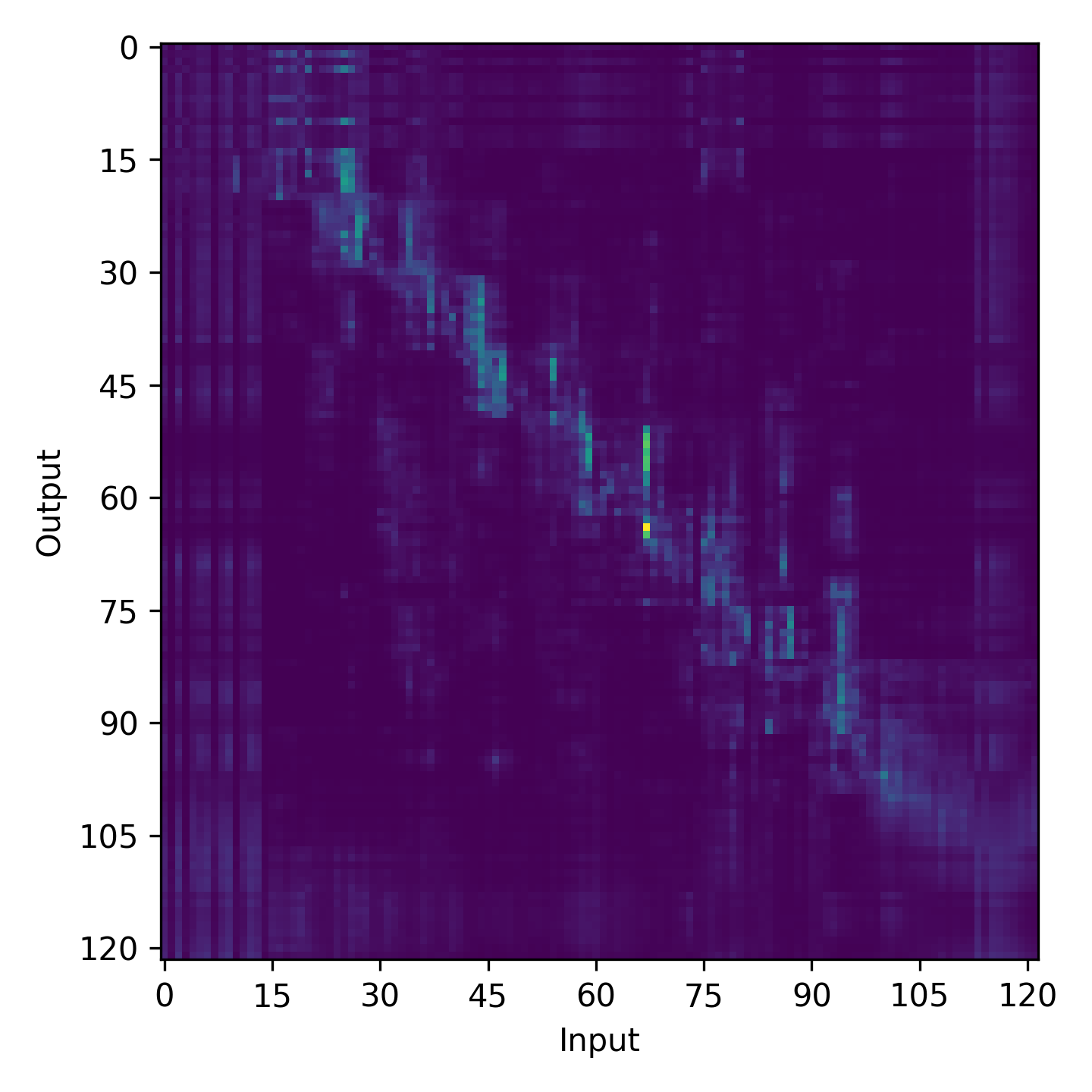}}
\centerline{(a)}
%\caption{fig1}
\end{minipage}%
\begin{minipage}[t]{0.33\linewidth}
\centering
\centerline{\includegraphics[width=1.1in]{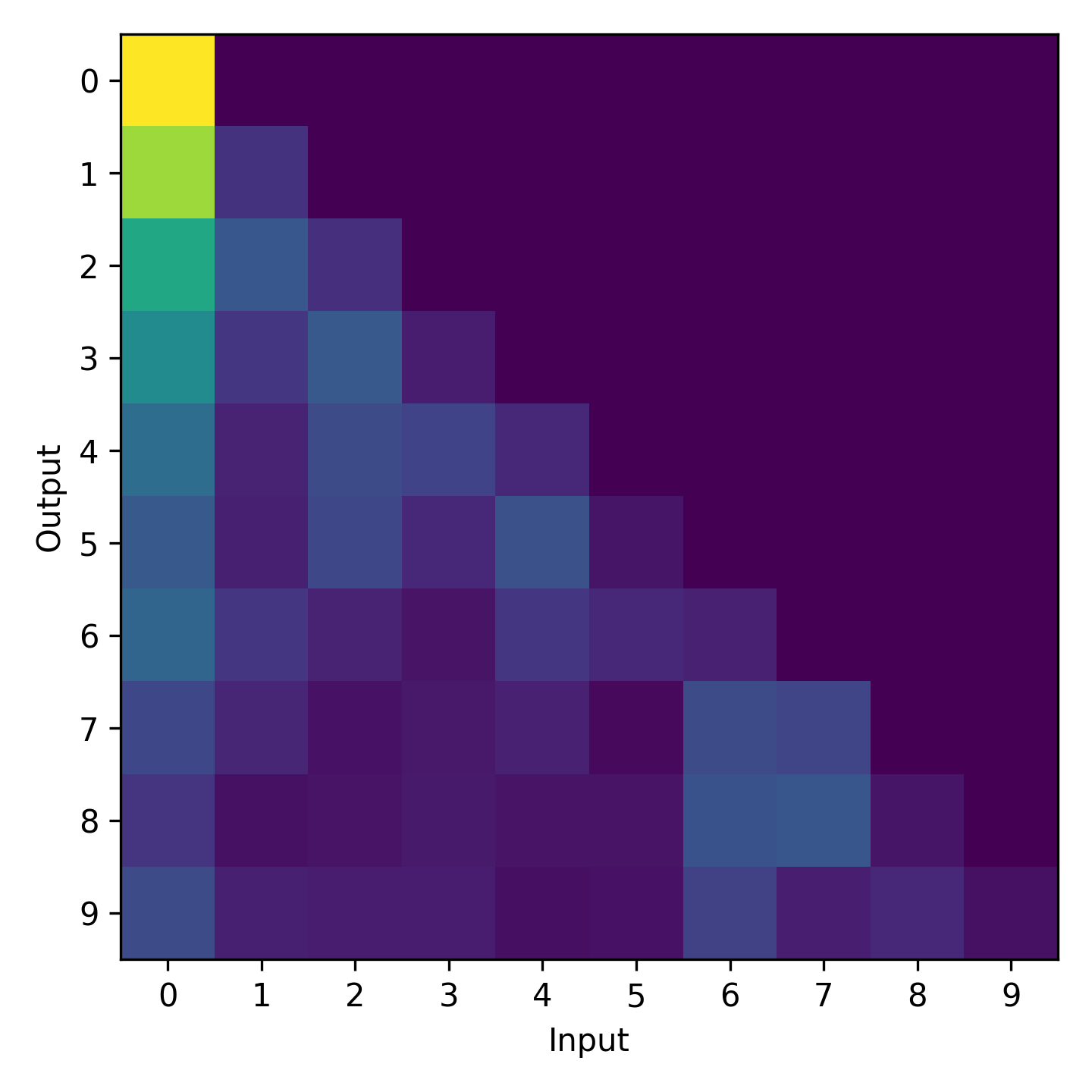}}
\centerline{(b)}
%\caption{fig2}
\end{minipage}%
\begin{minipage}[t]{0.33\linewidth}
\centering
\centerline{\includegraphics[width=1.1in]{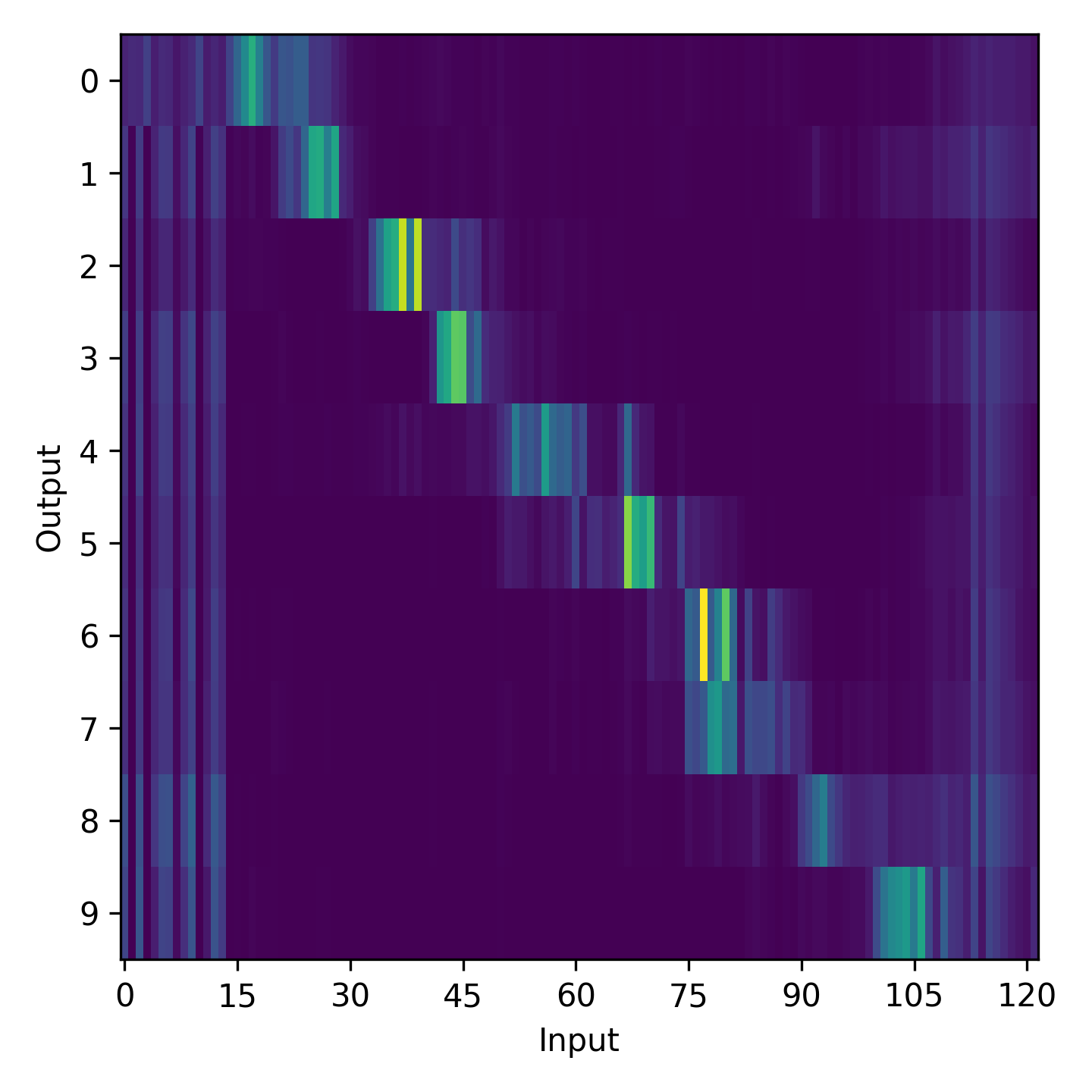}}
\centerline{(c)}
%\caption{fig3}
\end{minipage}%

\begin{minipage}[t]{0.33\linewidth}
\centering
\centerline{\includegraphics[width=1.1in]{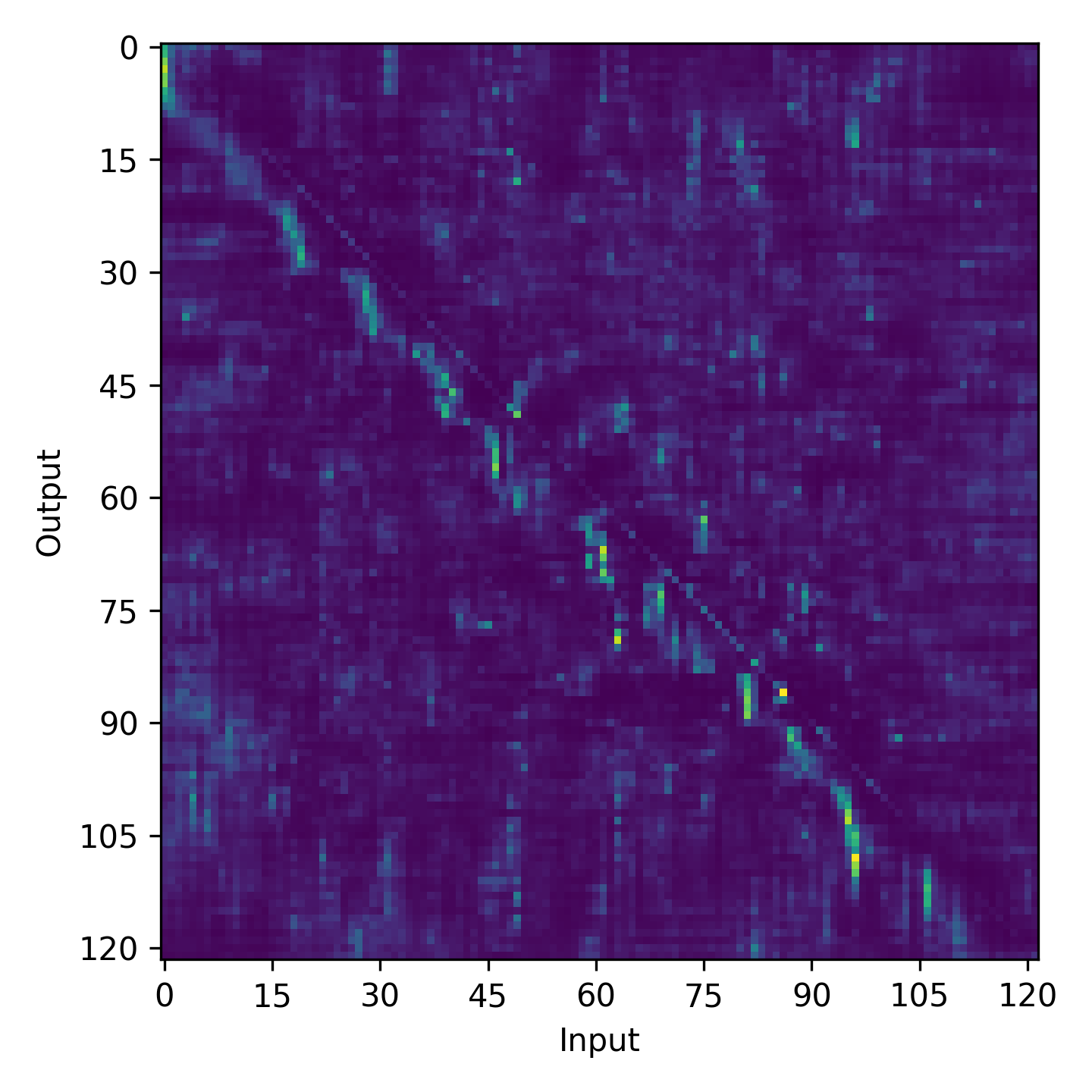}}
\centerline{(d)}
%\caption{fig1}
\end{minipage}%
\begin{minipage}[t]{0.33\linewidth}
\centering
\centerline{\includegraphics[width=1.1in]{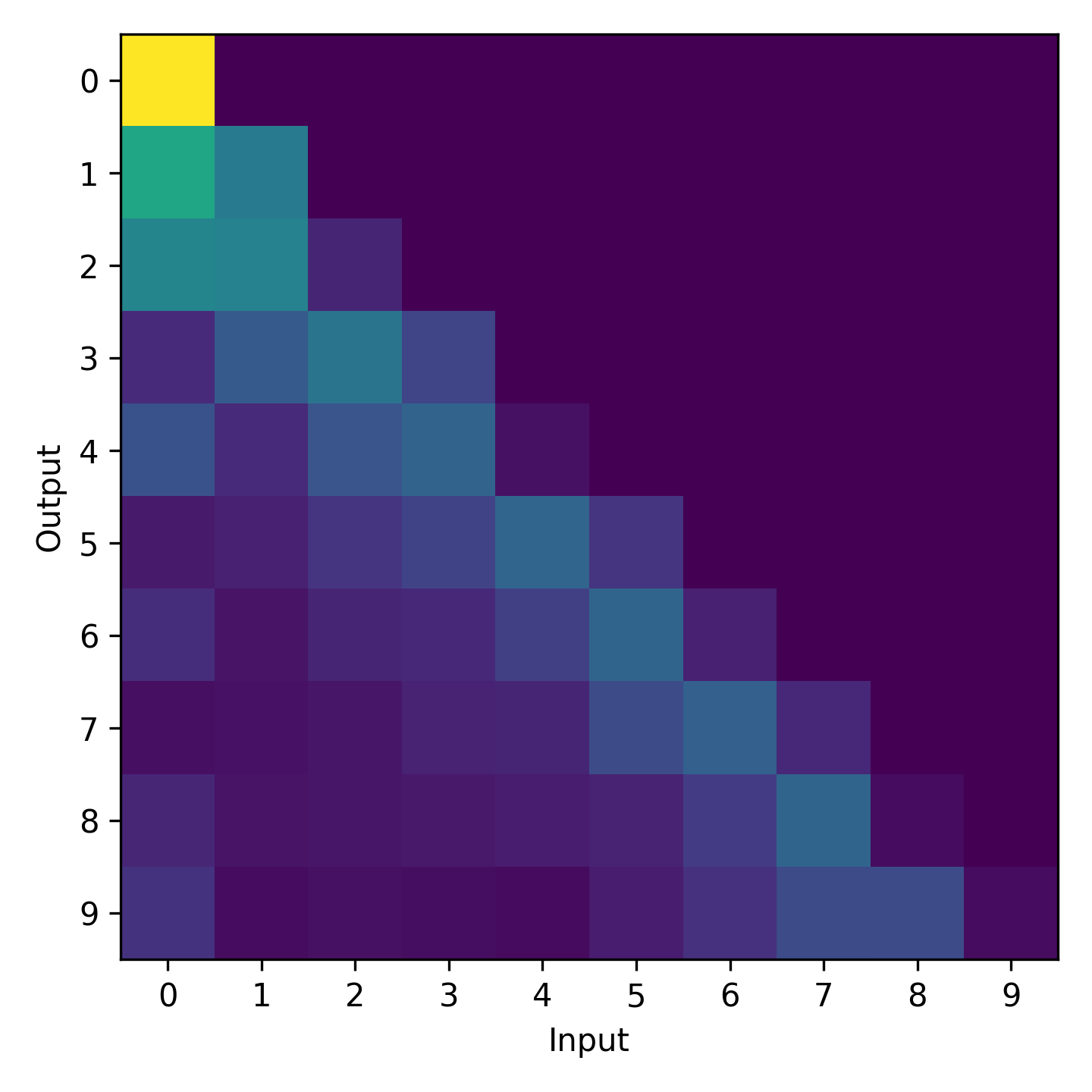}}
\centerline{(e)}
%\caption{fig2}
\end{minipage}%
\begin{minipage}[t]{0.33\linewidth}
\centering
\centerline{\includegraphics[width=1.1in]{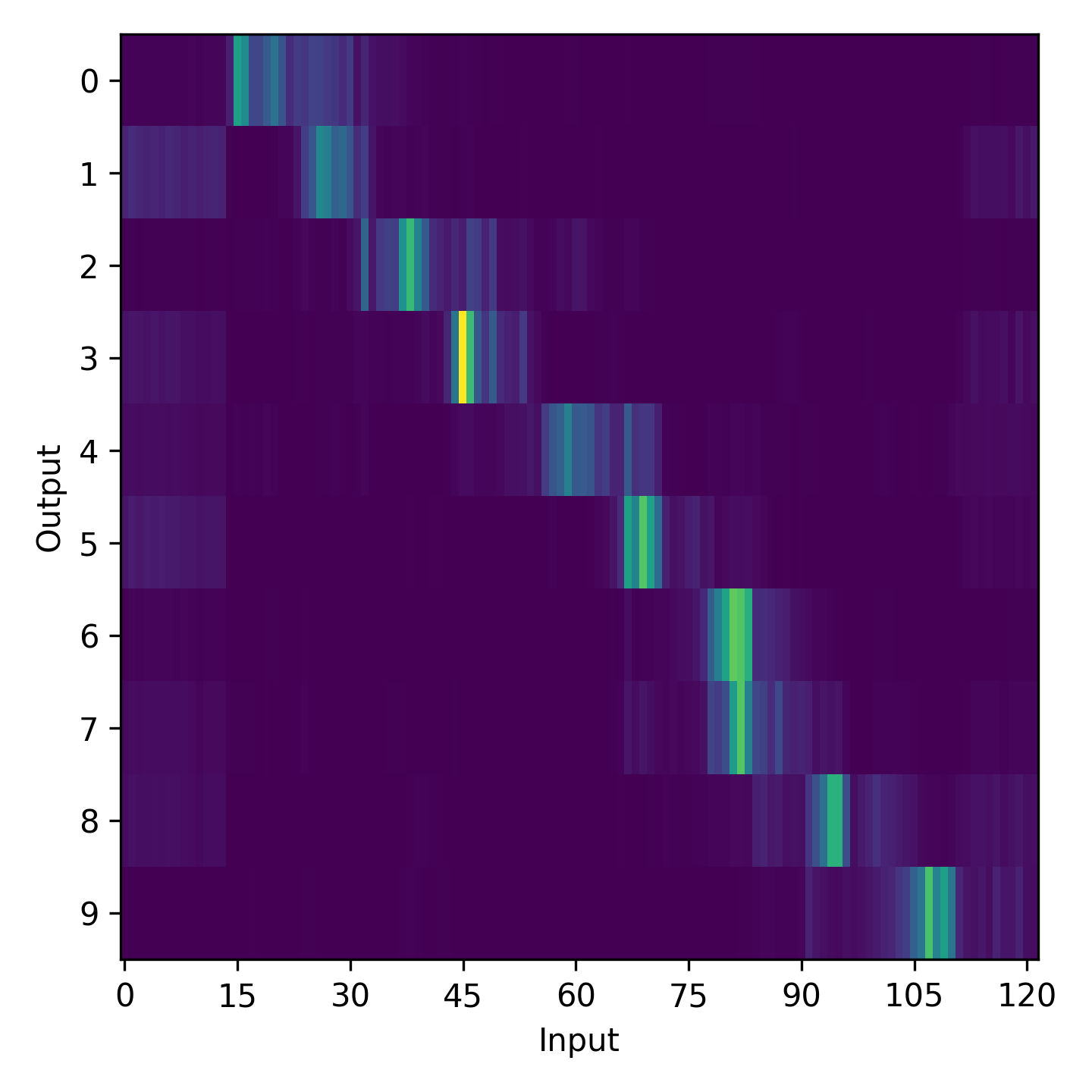}}
\centerline{(f)}
%\caption{fig3}
\end{minipage}%

\caption{Visualization of encoder, decoder and cross attention on both SAN-based (upper) and SSAN-based (lower) transformer model. For encoder-attention (a) and (d), x-axis and y-axis both refer to acoustic frames. For decoder-attention (b) and (e), x-axis and y-axis both refer to characters. For cross-attention (c) and (f), x-axis refers to acoustic frames and y-axis refer to characters. All attention figures are drawn for the utterance index BAC009S0725W0157 in the AISHELL-1 evaluation set. We use the last layer of encoder, decoder and cross-attention matrices and average for multi-heads to draw these figures. }
\label{fig:attention}
\end{figure}

We first validate our approach on the publicly available AISHELL-1 dataset. In order to verify whether the key, query, and value in self-attention can be formed by a simple FSMN memory block, we run a series of experiments with different model architectures. Results in Table~\ref{tab:diffstruc} show that SSAN-based transformer not only outperforms the SAN-based transformer but also has reduced model size. This conclusion is consistent for different architectures with different number of layers. Specifically, the SSAN-based transformer with encoder of 10 layers and decoder of 3 layers obtains 6.7\% relative improvement in CER and 21.7\% reduction in model parameters compared to SAN-based transformer with the same layer number setup. In Fig.~\ref{fig:attention}, we visualize the attention for a testing utterance in encoder, decoder and cross-attention for both SAN- and SSAN-based transformers. As the monotonic characteristics of speech, the energies are mainly concentrated along the diagonal. Clear diagonal means better attention alignment. The figures clearly demonstrate that SSAN can learn better attention alignment compared to SAN, especially for the decoder-attention and the cross-attention. We visualized 50 random-selected utterances and SSAN can achieve consistently better attention alignment.
    
	We further compare our model with the other published competitive models on the same AISHELL-1 task, including TDNN-LFMMI\cite{povey2016purely}, LAS\cite{shan2019component} and joint CTC-attention\cite{karita2019a} based transformer. Results in Table~\ref{tab:comparison} demonstrate that the performance of our SSAN-based transformer is close to the state-of-the-art joint CTC-attention model. Moreover, our model is trained using the CE-loss only and decoded without an external language model.

\subsection{1000-hour and 20,000-hour Tasks}
We further verify the effectiveness of the proposed SSAN-based transformer on the medium and large scale datasets. For the 1000-hour task, we use the best model architecture on AISHELL-1 for experimentation, which consists of 10 layers of SSAN-based encoder and 3 layers of SSAN-based decoder. Similar to the conclusion drawn from AISHELL-1, experimental results in Table~\ref{tab:1kh} demonstrate that SSAN is helpful to improve the performance with reduced model size on the 1000-hour dataset. Specifically, it can achieve 6.0\% relative improvement on the far-field test set and 20.4\% reduction in model parameters.  

  For large-scale 20,000-hour task, we use a big model (10 layers encoder, 6 layers decoder). Experimental results in Table~\ref{tab:2wh} show that, trained on a large dataset, the SSAN-based transformer still can bring 2.3\% relative improvement in CER on the far-field test set, while achieves comparable CER with SAN-based transformer on the near-field set. Such superior performance is achieved with 19.4\% reduction in model parameters. 
 	
\captionsetup[subfigure]{labelformat=empty}

\begin{table}[!htb]
\normalsize
\caption{Comparison of SAN and SSAN models on the 1000-hour Mandarin speech recognition task.}  
\centering  
\begin{tabular}{|l|c|c|c|}  
\hline
\multirow{2}{*}{Model}&\multirow{2}{*}{Param. (M)}&
\multicolumn{2}{c|}{CER (\%)} \\
\cline{3-4}
&&far-field & near-field\\ \hline
\hline  
SAN & 49 & 32.74 & 13.71 \\ \hline  
SSAN & 39 & \textbf{30.79} & \textbf{13.50} \\ \hline   
\end{tabular}  
\label{tab:1kh} 
\end {table}
\begin{table}[!htb]
\normalsize
\caption{Comparison of SAN and SSAN models on the 20,000-hour Mandarin speech recognition task.}  
\centering  
\begin{tabular}{|l|c|c|c|}  
\hline
\multirow{2}{*}{Model}&\multirow{2}{*}{Param. (M)}&
\multicolumn{2}{c|}{CER (\%)} \\
\cline{3-4}
&&far-field & near-field\\ \hline
\hline  
SAN & 62 & 22.36 & \textbf{7.84} \\ \hline  
SSAN & 50 & \textbf{21.84} & 7.91 \\ \hline   
\end{tabular}   
\label{tab:2wh} 
\end {table}

  Finally, comparing the results in Table~\ref{tab:1kh} and ~\ref{tab:2wh}, we find that SSAN consistently performs better on far-field test set than near-field test set. We believe that context information is more important for challenging far-field speech recognition, in which speech signal has lower quality due to signal degradation, room reverberation and noise interference. Our proposed FSMN-enhanced self-attention structure can better model long context, leading to substantial performance gain in far-field scenario. This conclusion is consistent with the experimental phenomena reported in~\cite{you2019dfsmn}, where augmented persistent memory can help to capture longer context information, resulting in better performance on far-field test sets.

% Below is an example of how to insert images. Delete the ``\vspace'' line,
% uncomment the preceding line ``\centerline...'' and replace ``imageX.ps''
% with a suitable PostScript file name.
% -------------------------------------------------------------------------

%\vfill
%\pagebreak

\section{Conclusions}

In this paper, we proposed a simplified self-attention network network (SSAN) layer by combining FSMN memory block for transformer ASR. We find that FSMN memory block can help the attention layer modeling longer context with substantial model parameter reduction. To make experimental results more convincing, we conducted a series of experiments on public AISHELL-1 corpus and internal industrial-level 1000- and 20,000-hour datasets. Results demonstrated the efficacy of our approach. As compared with the conventional transformer model, the SSAN-based transformer achieved improved performance on AISHELL-1 task and 1000-hour task and comparable performance on 20000-hour task. Impressively, the SSAN-based transformer reduced about 20\% of the model parameter.

%\section{REFERENCES}
%\label{sec:ref}

\bibliographystyle{IEEEbib}
\bibliography{strings,refs}

\end{document}